\documentclass[12pt]{amsart}
\usepackage[pdftex,bookmarks,colorlinks=true,linkcolor=blue,urlcolor=red]{hyperref}
\usepackage{amssymb}
\usepackage{amsfonts}
\usepackage{amsthm}
\usepackage{amsmath}
\usepackage{mathtext}
\usepackage{latexsym}
\numberwithin{equation}{section}
\newcommand{\im}{{\rm Im}\,}
\newcommand{\re}{{\rm Re}\,}
\newcommand{\C}{{\mathbb C}}
\newcommand{\R}{{\mathbb R}}
\newcommand{\Z}{{\mathbb Z}}
\newcommand{\N}{{\mathbb N}}
\newcommand{\Q}{{\mathbb Q}} 
\newcommand{\vers}{\operatornamewithlimits{\to}}

\newcommand{\D}{\displaystyle}

\def\im{\operatorname{Im}}
\def\re{\operatorname{Re}}
\theoremstyle{plain}
\newtheorem{Th}{Theorem}
\newtheorem{Le}{Lemma}

\newtheorem{Cor}{Corollary}
\theoremstyle{definition}

\title{Stark-Wannier ladders and cubic exponential sums}
\author{Alexander Fedotov} \author{Fr{\'e}d{\'e}ric Klopp}
\address[Alexander Fedotov]{St. Petersburg State University, 
7/9 Universitetskaya nab., St.Petersburg, 199034, Russia}
\email{\href{mailto: a.fedotov@spbu.ru}{a.fedotov@spbu.ru}}
\address[Fr{\'e}d{\'e}ric Klopp]{ \vskip.1cm Sorbonne Universit{\'e}s,
  UPMC Univ. Paris 06, UMR 7586, IMJ-PRG, F-75005, Paris, France
  \vskip.1cm Univ. Paris Diderot, Sorbonne Paris Cit{\'e}, UMR 7586,
  IMJ-PRG, F-75205 Paris, France
  \vskip.1cm CNRS, UMR 7586, IMJ-PRG, F-75005, Paris, France}
\email{\href{mailto:frederic.klopp@imj-prg.fr}{frederic.klopp@imj-prg.fr}}
\date{} 
\thanks{The present work was supported by the Russian foundation of
  basic research under grant 14-01-00760-a. A.F. acknowledges support
  by the Fondation Sciences Math{\'e}matiques de Paris. F.K. acknowledges
  support by the Chebyshev Laboratory and the French Embassy in Russia
  through the Chaire Lam{\'e}. The authors would like to thank the Isaac
  Newton Institute for Mathematical Sciences for its hospitality
  during the programme ``Periodic and Ergodic Spectral Problems''
  supported by EPSRC Grant Number EP/K032208/1. F.K. also acknowledges
  support from the Simons Foundation during his stay at the INI.}
\begin{document}
\maketitle
%
%
\noindent On $ L^2(\R)$, we consider the Schr{\"o}dinger operator
\begin{equation}\label{eq:operator}
  H_\epsilon=-\frac{\partial^2}{\partial x^2}+v(x) -\epsilon x,
\end{equation}
where $v$ is a real analytic $1$-periodic function and $\epsilon$ is a
positive constant. This operator is a model to study a Bloch electron
in a constant electric field (\cite{Av:1982}). The parameter
$\epsilon$ is proportional to the electric field.  The
operator~\eqref{eq:operator} was studied both by physicists (see,
e.g., the review~\cite{Gl-Kz-Ksch:2002}) and by mathematicians (see,
e.g.,~\cite{MR3390534}). Its spectrum is absolutely continuous and
fills the real axis. One of main features of $H_\epsilon$ is the
existence of Stark-Wannier ladders. These are $\epsilon$-periodic
sequences of resonances, which are poles of the analytic continuation
of the resolvent kernel in the lower half plane through the spectrum
(see, e.g.,~\cite{Bu-Dm:1990}).  Most of the mathematical work studied
the case of small $\epsilon$ (see, e.g.,~\cite{MR3390534,Bu-Gr:1998}
and references therein). When $\epsilon$ is small, there are ladders
exponentially close to the real axis. Actually, only the case of
finite gap potentials $v$ was relatively well understood. For these
potentials, there is only a finite number of ladders exponentially
close to the real axis.  It was further noticed that the ladders
non-trivially ``interact'' as $\epsilon$ changes, and conjectured that
the behavior of the resonances strongly depends on number theoretical
properties of $\epsilon$ (see, e.g.,~\cite{Av:1982}).
\par In the present note, we only consider the periodic potential
$v(x)=2\cos(2\pi x)$ and study the reflection coefficient $r(E)$ of
the Stark-Wannier operator~\eqref{eq:operator} in the lower half of
the complex plane of the spectral parameter $E$. The resonances are
the poles of the reflection coefficient. We show that, as ${\rm
  Im}\,E\to-\infty$, the function $\D E\mapsto\frac1{r(E)}$ can be
asymptotically described in terms of a regularized cubic exponential
sum that is a close relative of the cubic exponential sums often
encountered in analytic number theory. This explains the dependence of
the reflection coefficient on the arithmetic nature of $\epsilon$. For
$\frac{\pi^2}{3\epsilon}\in\Q$, we describe the asymptotics of the
Stark-Wannier ladders situated far from the real axis.
\par Let us recall the definition of the reflection coefficient
for~\eqref{eq:operator} following~\cite{Bu-Dm:1990}. Consider the
equation
\begin{equation}
  \label{equation}
  -\psi''(x)+(v(x) -\epsilon x)\psi(x)=E\psi(x),\quad x\in\C,
\end{equation}
For the sake of simplicity, assume that the potential $v$ is entire.
Assume also $\D\int_0^1v(x)\,dx=0$. For any $E\in\C$, there are unique solutions
$\psi_\pm$ to~\eqref{equation} that admit the asymptotic
representations
\begin{equation}
  \label{eq:1}
  \begin{aligned}
  \psi_-(x,E)&={\textstyle\frac1{\sqrt[4]{-\epsilon x-E}}}
  e^{-\int_x^{-E/\epsilon} \sqrt{-\epsilon t-E}\,dt +o(1)},\quad
  x\to-\infty,\\
  \psi_+(x,E)&={\textstyle\frac1{\sqrt[4]{\epsilon x+E}}}
  e^{i\int_{-E/\epsilon}^x \sqrt{\epsilon t+E}dt +o(1)},\quad
  x\to+\infty,    
  \end{aligned}
\end{equation}
where the determinations of $\sqrt{\cdot}$ and $\sqrt[4]{\cdot}$ are
analytic in $\C\setminus\R_-$ and positive along $\R_+$. Consider also
the solution $\psi_+^*(x,E)=\overline{\psi_+(\bar x,\bar E)}$.  The
solutions $\psi_+$ and $\psi_+^*$ being linearly independent, one has
\begin{equation}
  \label{eq:M}
  \psi_-(x,E)=w(E)\psi_+^*(x,E)+w^*(E)\psi_+(x,E),\quad x\in\R,
\end{equation}
where the coefficient $w(E)$ is independent of $x$ and the function
$E\mapsto w(E)$ is entire.  The ratio $r(E)=w^*(E)/w(E)$ is {\it the
  reflection coefficient}. It is an $\epsilon$-periodic meromorphic
function of $E$. The reflection coefficient is analytic in $\C_+$,
and, for $E\in \R$, one has $|r(E)|=1$. The poles of $r$ are the
resonances of $H_\epsilon$.
\par Let us now state the first of our results. Represent $1/r$ by its
Fourier series $\D 1/r(E)=\sum_{m\in\Z} e^{2\pi n i E/\epsilon}p(m)$ for 
$\im E\le 0$. Let $a(\epsilon)=\sqrt{\frac{2}{\epsilon}}\,\pi e^{i\pi/4}$. 
One has
\begin{Th}
  \label{thr:1}
  Let $v(x)=2\cos(2\pi x)$.  Then, as $m\to\infty$,
  \begin{equation}
    \label{pm}
    p(m)=a(\epsilon)\,\sqrt{m}\,
    e^{-2\pi i \omega m^3-2m\log\,(2\pi m/e)+\delta(m)},\quad {\textstyle 
      \omega=\left\{\frac{\pi^2}{3\epsilon}\right\}}, 
  \end{equation}
  where, for $x$ real, $\{x\}$ denotes the fractional part of $x$, and
  $\delta(m)=O(\log^2m/m)$. This estimate is locally uniform in
  $\epsilon>0$.
\end{Th}
\noindent Clearly, the asymptotic behavior of $1/r(E)$ as $\im
E\to-\infty$ is determined by the Fourier series terms with large
positive $m$, and so, roughly,
\begin{equation}
  \label{r:cubic}
  \frac1{r(E)}\approx a(\epsilon)\, {\mathcal P}(E/\epsilon),\quad
  {\mathcal P}(s)=\sum_{m\ge 1} \sqrt{m}\,
  e^{-2\pi i \omega m^3-2m\log\,(2\pi m/e)+2\pi i m s}.
\end{equation}
It is worth to compare the function $\mathcal P$ with the cubic
exponential sums $\D\sum_{n=1}^N e^{-2\pi i \omega n^3}$.  Such sums
were extensively studied in analytic number theory, see,
e.g.,~\cite{Da:2005}. They were proved to depend strongly on the
arithmetic nature of $\omega$.  This appears to be true in our case
too. We have
\begin{Th}
  \label{r:as}
  Let $v(x)=2\cos(2\pi x)$. Assume that $\omega\in\Q$ and represent it
  in the form $\omega=\frac{p}{q}$, where $0\le p<q$ are co-prime
  integers. If $p=0$, we take $q=1$.  For $\xi\in\R$, we set
  $I_q(\xi):=\{m\in\Z:\,|\xi-\frac{m}q|\le1/2\}$.  As $\im
  E\to-\infty$, one has
  \begin{equation}\label{eq:r:as}
    r^{-1}(E)={\textstyle \frac{b(\epsilon)\,\rho}q}
    \sum_{m\in I_q(\xi)}S_{q}(p,m)e^{\rho e^{i\pi(\xi-m/q)}+i\pi(\xi-m/q)
      +O(\log^2\rho/\rho)}+e^{O(\frac{\rho}{\ln\rho})},
  \end{equation}
  where $b(\epsilon)=\pi^{\frac32} e^{i\pi/4}/\sqrt{2\epsilon}$, \
  $\xi={\rm Re}\,E/\epsilon$, \ $\rho=e^{-\pi\,{\rm Im}\,E/\epsilon}$,
  and
  \begin{equation*}
    S_{q}(p,m)=\sum_{l=0}^{q-1} e^{-2\pi i\,\frac{pl^3-ml}q}.
  \end{equation*}
  The error estimates are locally uniform in $\epsilon>0$.
\end{Th}
\noindent Let us discuss this result. First, assume that $\omega=0$.
By Theorem~\ref{r:as},
\begin{equation}\label{q:1}
  (b(\epsilon)r(E))^{-1}=\sqrt{z}\,
  e^{\sqrt{z}+O(\frac{\ln^2z}{\sqrt{z}})}+
  e^{O(\frac{\sqrt{z}}{\ln z})},
  \quad z=e^{2i\pi E/\epsilon},
\end{equation}
where the determination of $\sqrt{\cdot}$ is analytic in
$\C\setminus\R_-$ and positive along $\R_+$. Recall that $1/r$ is
$\epsilon$-periodic.  Let $B_\epsilon=\{E\in\C:\;\im E\le 0, \,0\le
\re E\le \epsilon\}$.  Representation~\eqref{q:1} implies
\begin{Cor}
  \label{cor:1}
  Assume $\omega=0$. The resonances located in $B_\epsilon$ have the
  following properties :
  \begin{itemize}
  \item for sufficiently large $y>0$, the resonances with $\im
    E<-\epsilon y$ are located in the domain $|\re E-\epsilon/2|\le
    C\epsilon^2/|\im E|$, where $C>0$ is a constant;
  \item let $n(y)$ be the number of resonances in the rectangle
    $\D[0,\varepsilon]-i\left[0,\varepsilon y\right]$; then, one has
    \begin{equation*}
      n(y)=\frac1{\pi}{}\,e^{\pi y+o(1)}\quad\text{as}\quad
      y\to\infty. 
    \end{equation*}
  \end{itemize}
\end{Cor}
\noindent The first statement immediately follows from Theorem~\ref{r:as}; 
to prove the second one has to use Jensen formula and Levin lower bounds 
for the absolute values of entire functions, see, e.g.,~\cite{MR589888}. \\
When $\omega=0$, it is difficult to obtain the asymptotics of the
resonances as, in a neighborhood of the line $\re E/\epsilon=1/2\mod
1$, they are determined by the first Fourier coefficients of $1/r$,
i.e., by $p(m)$ with $m=1,2,3\dots$. Hence, the problem is not
asymptotic in nature.
\par If $\omega\ne 0$, then the description of the resonances is
determined by the values of $S_q(p,m)$ for $m=1,2,\dots q-1$ (the map 
$m\to S_q(p,m)$ is $q$-periodic). The $S_q(p,m)$ are {\it cubic complete 
rational exponential sums}, see, e.g.,~\cite{Ko}. One easily checks
\begin{Le}
  \label{S-squares}
  For any $q\in\N$, \ $\D\sum_{m=0}^{q-1}|S_q(p,m)|^2=q^2$.
\end{Le}
\noindent This implies that, for $q\ge 1$, there is at least one
integer $0\le m_0<q-1$ such that $S_q(p,m_0)\ne0$.

If $S_q(p,m)$ is non zero for only one $0\le m_0<q$ (this happens, for
example, for $q=2,3,6$), then one can characterize the resonances as
when $\omega=0$. Now, they live near the lines $\{\re E/\epsilon=
m_0/q+1/2+n\}$, $n\in\Z$.
\par For large $q$, there are actually many non-zero values
$S_q(p,m)$:
\begin{Le}
  There exists a constant $C>0$ such that, for any co-prime $q>p>0$,
  one has $\#\{0\le m<q:\, S_q(p,m)\ne 0\}\ge C q^{\frac23}$.
\end{Le}
\noindent This statement follows from Lemma~\ref{S-squares} and the
well-known upper bound for general complete rational exponential
sums of Hua (\cite{Ko}).
\par In general, the behavior of $m\mapsto S_q(p,m)$ is nontrivial; it
is known to depend strongly on the prime factorization of $q$.
Computer calculations lead to the following conjecture: {\it if $q$ is
  prime, $0<p<q$, and $0<m<q$, then $S_q(p,m)\ne 0$}.

If $S_q(p,m)$ is non zero for at least two values of $m$ such that
$0\le m<q$, then, using~\eqref{pm}, one can describe asymptotically
all the resonances with sufficiently ne\-gative imaginary part. One has
\begin{Cor}
  Assume that, for some integers $m_1<m_2$ such that $m_2-m_1<q$, one
  has $S_q(p,m_1)\ne0$, $S_q(p,m_2)\ne0$, and $S_q(p,m)=0$ for all
  $m_1<m<m_2$.  Then, for sufficiently large $y>0$, in the vertical
  half-strip
  \begin{equation*}
    \left\{E\in\C:\; -\im E\ge \epsilon y,\;\;\frac{m_1}q\le \frac{\re
        E}\epsilon\le\frac{m_2}q\right\} ,
  \end{equation*}
  there are resonances, and they are described by the asymptotic
  formulas: 
  \begin{equation}\label{eq:res}
    \frac{E}\epsilon=
    -i\left(\frac{\ln(\pi k)}{\pi}-\ln\sin\frac{\pi(m_2-m_1)}q\right)+
    \frac{m_2+m_1}q+o(1),\quad k\in \N,
  \end{equation}
  where $\D o(1)\vers_{k\to+\infty}0$.
\end{Cor}
\noindent This statement easily follows from Theorem~\ref{r:as}.
\par Finally, let us describe very briefly the ideas leading to
Theorems~\ref{thr:1} and~\ref{r:as}.  Buslaev's solutions $\psi_\pm$
used to define the reflection coefficient (see~\eqref{eq:1}) are
entire functions of $x$ and $E$; they satisfy the relations
$\psi_\pm(x+1,E)=\psi_\pm(x,E+\epsilon)$. It appears that the analytic
properties of such solutions can naturally be described in terms of a
system of two first order difference equations on the complex plane
(see, for example,~\cite{Fe-Kl:2002}). To get the asymptotics of the
Fourier coefficients of the reflection coefficient, we study the
solutions of this system far from the origin. The idea leading from
Theorem~\ref{thr:1} to Theorem~\ref{r:as} is analogous to one used to
study the behavior of the exponential sums $\D\sum_{n=1}^N e^{-2\pi i
  \omega n^3}$ with $\omega\in\Q$ for large $N$,
see~\cite{Da:2005}. However, to use it successfully, one has to carry
out a non trivial analysis of properties of the error term
in~\eqref{pm}.
%
%
\bibliographystyle{plain}

%

\end{document}